\documentclass[iop,apj,12pt,numberedappendix]{emulateapj}
\usepackage{times}
\usepackage[normalem]{ulem}
\usepackage{graphicx}
\usepackage{natbib}
\usepackage{amsfonts}
\usepackage{amsmath}
\usepackage{multirow}
\usepackage{enumerate}
\usepackage{comment}
\usepackage{float}
\usepackage{verbatim}
\usepackage[usenames,dvipsnames]{xcolor}
\usepackage[plainpages=false, colorlinks=true, anchorcolor=blue, linkcolor=blue, citecolor=blue, bookmarks=false]{hyperref}
\newcommand{\be}{\begin{eqnarray}}
\newcommand{\ee}{\end{eqnarray}}

\newcommand{\shortauth}{Morozova et al.}
\newcommand{\slugcom}{Submitted for publication in The Astrophysical Journal}
\slugcomment{\slugcom}

\lefthead{\sc \footnotesize \slugcom \hfill \shortauth}
\righthead{\sc \footnotesize \slugcom \hfill \shortauth}

\begin{document}

\title{Unifying Type II Supernova Light Curves with Dense Circumstellar Material}

\author{Viktoriya Morozova\altaffilmark{1,2}}
\author{Anthony L. Piro\altaffilmark{3}}
\author{Stefano Valenti\altaffilmark{4}}
\altaffiltext{1}{Department of Astrophysical Sciences,
  Princeton University, Princeton, NJ 08544, USA, 
  vsg@astro.princeton.edu}
\altaffiltext{2}{TAPIR, Walter Burke Institute for Theoretical Physics, MC 350-17,
  California Institute of Technology, Pasadena, CA 91125, USA}
\altaffiltext{3}{Carnegie Observatories, 813 Santa Barbara Street, Pasadena, CA 91101, USA}
\altaffiltext{4}{Department of Physics, University of California, Davis, CA 95616, USA}

\begin{abstract}

A longstanding problem in the study of supernovae (SNe) has been the
relationship between the Type~IIP and Type~IIL subclasses. Whether they come from
distinct progenitors or they are from similar stars with some property that
smoothly transitions from one class to another
has been the subject of much debate. Here we show using one-dimensional
radiation-hydrodynamic SN models that the multi-band light curves of SNe IIL are
well fit by ordinary red supergiants surrounded by dense circumstellar
material (CSM). The inferred extent of this material, coupled with a typical wind
velocity of $\sim10-100\,{\rm km\,s^{-1}}$, suggests enhanced activity by these
stars during the last $\sim\,$months to $\sim\,$years of their lives, which may be
connected with advanced stages of nuclear burning. Furthermore, we find that even for
more plateau-like SNe that dense CSM provides a better fit to
the first $\sim20\,$days of their light curves, indicating that the presence
of such material may be more widespread than previously appreciated. Here
we choose to model the CSM with a wind-like density profile, but it is unclear whether
this just generally represents some other mass distribution, such as a recent
mass ejection, thick disk, or even inflated envelope material. Better understanding
the exact geometry and density distribution of this material will be an important
question for future studies.

\end{abstract}

	
  
\section{Introduction}

Hydrogen-rich supernovae (SNe) have traditionally been divided
into Type IIP (plateau) and Type IIL (linear) subclasses based on the
shape of their light curves during the first few weeks \citep{barbon:79}.
Beyond just their light curve morphology, these subclasses have other
distinguishing features, such as SNe~IIL are on average more luminous than
SNe~IIP by $\sim\,$1.5 mag \citep{patat:93,patat:94,anderson:14,faran:14a,sanders:15},
SNe~IIL tend to have redder continua and
 higher oxygen to hydrogen ratio as compared to ordinary SNe~IIP
\citep{faran:14a}, SNe~IIL exhibit higher expansion velocities at early times \citep{faran:14a},
and they have less pronounced P-Cygni H$\alpha$ profiles \citep{gutierrez:14}.

These differences have inspired a long debate on whether there is
a physical process that smoothly transitions between Type IIP and IIL or
whether there is a specific mechanism that creates this distinction more
abruptly. Although there have been claims of distinct populations \citep{arcavi:12,faran:14b,faran:14a},
support for the more continuous case has increased as larger compilations
 by \citet{anderson:14} and \citet{sanders:15} showed a
more continuous range of early light curve slopes. Following this, \citet{valenti:15} importantly
demonstrated that if one
simply follows a SN IIL long enough, its light curve will drop at $\sim100\,{\rm days}$,
just like a normal SN IIP (previous SNe IIL studies rarely followed the light curve
beyond $\sim80\,{\rm days}$ from discovery). This implies that
Type~IIL and Type~IIP SNe may share the
same basic progenitor, and whatever is creating the Type IIL distinction
must be contributing something {\em above} the fairly normal underlying red supergiant (RSG).

At the same time, there has been increasing evidence of SNe
interacting with dense circumstellar material (CSM) that requires strong
mass loss shortly before core collapse
\citep[see][and references therein]{smith:14}. This can manifests itself in
narrow optical emission lines \citep{filippenko:97,
pastorello:08,kiewe:12,taddia:13},
X-ray or radio emission \citep{campana:06,corsi:14}, or a rapid
rise at ultraviolet wavelengths 
\citep{ofek:10,gezari:15,tanaka:16,ganot:16}.
In the most extreme cases, there are the
super-luminous SNe and SNe IIn events that can require
$\sim10\,M_\odot$ or more ejected in the last few years of a massive
star's life \citep{smith:07b,smith:07,woosley:07b,vanmarle:10,smith:11b,ofek:13}.
Nevertheless, it has also become clear that many other SNe
have fleeting signs of CSM interaction where SNe IIn spectral features
are seen within a few days of explosion \citep{gal-yam:14,smith:15,khazov:16}.
This indicates that smaller, but still dramatic mass loss may be more widespread.
In the particular case of PTF11iqb, it transitioned from showing Type~IIn-like features to
a Type~IIL light curve before becoming more IIP-like and finally showing
IIn features again \citep{smith:15}. This suggests an even closer relationship between
these SN types and the CSM properties, and that in many cases we might just
lack the temporal coverage (especially at early and late times) needed to identify the CSM's impact.

Motivated by these developments, we undertake a theoretical
study on the affect of dense CSM around RSGs on SN light curves,
and then conduct detailed comparisons with observed SNe IIP and IIL.
We begin in Section~\ref{impact} by summarizing our numerical
methods and presenting a series of simulations to survey the range
of ways a dense CSM will alter light curves. In Section~\ref{overview},
we provide a brief overview of  SNe~2013ej, 2013by and 2013fs,
three SNe for which we then conduct detailed,
multi-band fits in Section~\ref{fitting}.
We discuss the application of our study to the problem of 
diversity between the SNe IIP and IIL in Section~\ref{discussion},
and discuss the implications for the nature of the mass loss inferred from our fits. 
Finally, we 
summarize our conclusions in Section~\ref{conclusion}.


\section{Impact of a Dense Wind on Light Curves}
\label{impact}

We begin by outlining our numerical setup and
presenting a series of simulations to explore
the impact of a dense wind on SN light curves. This will
help provide some guidance on what light curve features
can be affected for our later comparison to observations.

\subsection{Numerical Setup}

Throughout this work we use
the non-rotating solar-metallicity RSG models from
the stellar evolution code \texttt{KEPLER}
\citep{weaver:78,woosley:07,woosley:15,sukhbold:14,sukhbold:16}.
Extending above these models we add a dense CSM,
for which we assume a steady-state wind with
a density profile
\be
	\rho(r) = \frac{\dot{M}}{4\pi r^2 v_{\rm wind}} = \frac{K}{r^2},
\ee
where $\dot{M}$ is the wind mass loss rate and $v_{\rm wind}$ is the wind
velocity. In general, we infer $\dot{M}$ from
our models based on the $K$ we are using and the
expected $v_{\rm wind}$. This density profile extends out to a radius
$R_{\rm ext}$ where we abruptly set the density to zero. This provides us
with a useful parameterization for exploring the properties of the CSM
(with the main variables being $K$ and $R_{\rm ext}$).
It is possible that the CSM may actually be in some other mass distribution,
and this wind we consider is just an approximation.
We discuss this possibility further in Section~\ref{discussion}.

The impact of the dense wind has been 
investigated in a large number of works \citep{smith:07,chugai:07,
ofek:10,chevalier:11,moriya:11}. The main difference between our
work and these previous studies is that
 we focus on considerably higher
mass losses ($\dot{M}$ in the range of $0.02-15\,M_{\odot}\,{\rm yr}^{-1}$)
and small external radii of the wind 
$900\,R_{\odot}<R_{\rm ext}<2700\,R_{\odot}$. In contrast, for example,
\citet{moriya:11} considers larger
radii ($\sim 10^4\,R_{\odot}$) and wind mass loss rates in the range
of $10^{-4}-10^{-2}\,M_{\odot}\,{\rm yr}^{-1}$. The closest analog to our work is in
fact probably the study by \citet{nagy:16}, who consider a two-component
model for fitting Type IIP light curves. In particular, both
that study and our work here attempt to fit SN~2013ej,
and we compare these results below.

These models are then exploded with our open-source numerical
code \texttt{SNEC} \citep{morozova:15}. We assume that the inner 
$1.4\,M_{\odot}$ of the models form a neutron star and excise this 
region before the explosion. We use a thermal bomb mechanism for the
explosion, adding the energy of the bomb to the internal energy in
the inner $0.02\,M_{\odot}$ of the model for a duration of $1\,{\rm s}$.
The compositional profiles are smoothed using ``boxcar'' approach
with the same parameters as in \citet{morozova:15}, and the same
values for the opacity floor are adopted. The equation of state includes 
contributions from ions, electrons and radiation, with the degeneracy
effects taken into account as in \citet{paczynski:83}. 
We trace the ionization fractions of
hydrogen and helium solving the Saha equations in the non-degenerate
approximation as proposed in \citet{zaghloul:00}. The numerical grid consists
of 1000 cells and is identical to the one used in \citet{morozova:15,morozova:16}.

We include the velocity of the wind in our models set to 
$10\,{\rm km}\,{\rm s}^{-1}$, as in \citet{moriya:11}. 
Note that the unshocked wind velocities measured
from Type IIn SNe are
higher than that and vary in the range $10^2-10^3\,{\rm km}\,{\rm s}^{-1}$
\citep[see][]{kiewe:12}.
The escape velocities of the RSG model that we use vary in the range
$75-92\,{\rm km}\,{\rm s}^{-1}$. The exact choice of
the wind velocity does not matter in detail though
because \citet{moriya:11}
find that the wind velocity has little impact on the final light curve,
which we confirm by comparing simulations performed with
wind velocities $10$ and $10^2\,{\rm km}\,{\rm s}^{-1}$.

\subsection{Parameter Survey Results}

\begin{figure}
  \centering
  \includegraphics[width=0.475\textwidth]{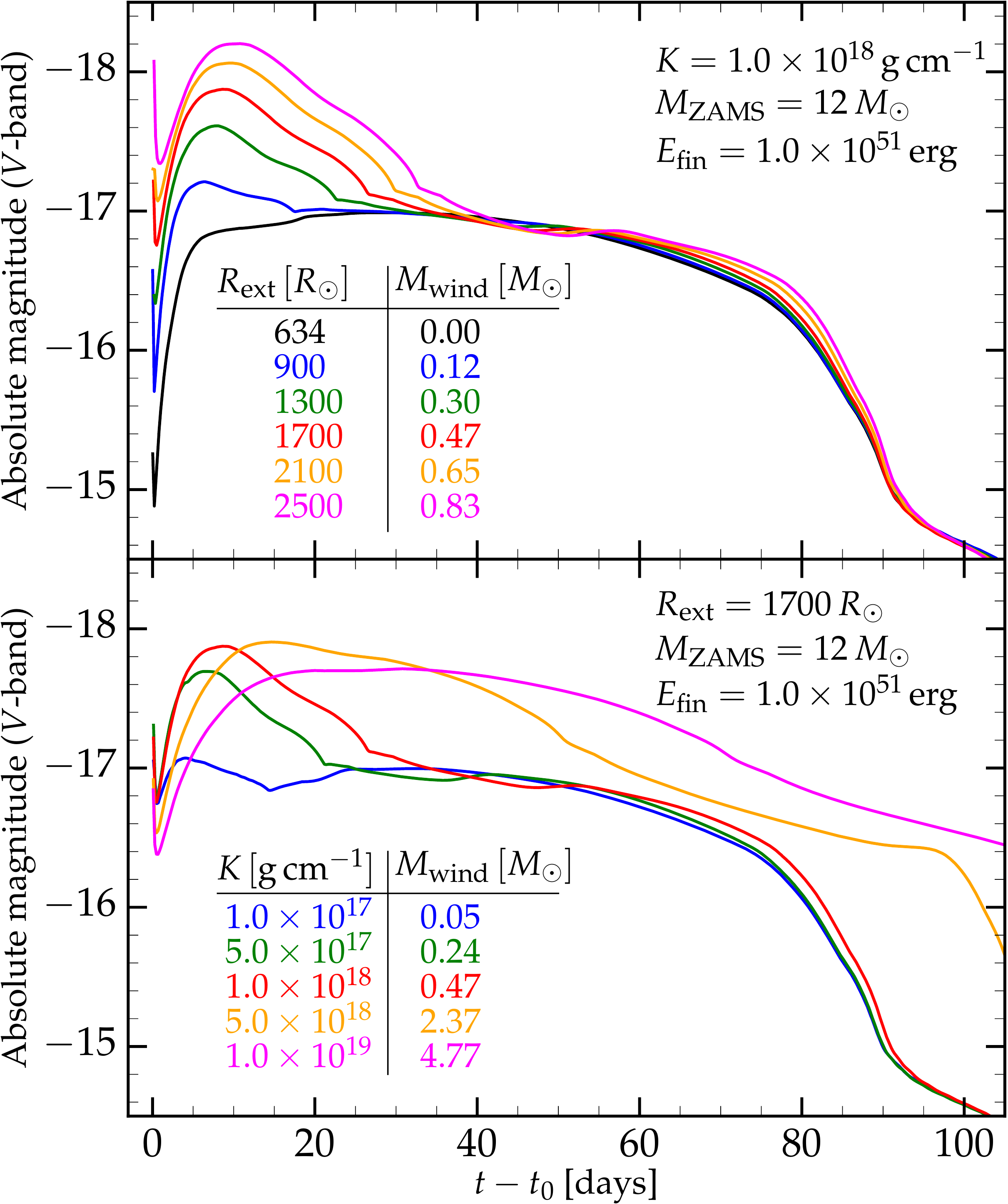}
  \caption{Top panel: $V$-band light curves of 
  $M_{\rm ZAMS} = 12\,M_{\odot}$ RSG model supplemented by
  a wind with 
  $K=1.0\times10^{18}\,{\rm g}\,{\rm cm}^{-1}$ and
  different values of the external radius $R_{\rm ext}$. 
  Bottom panel: $V$-band light curves of 
  the same model supplemented by
  a wind with a fixed external radius $R_{\rm ext}=1700\,R_{\odot}$
  and different values of the density parameter $K$.} 
  \label{fig:different}
\end{figure}

For our initial study of the CSM impact, we stitch the dense wind to a RSG model
with a zero-age main-sequence (ZAMS) mass of
$M_{\rm ZAMS} = 12\,M_{\odot}$ for the different values of parameters $R_{\rm ext}$
and $K$. The asymptotic energy of each explosion is 
$E_{\rm fin}=1.0\times10^{51}\,{\rm erg}$,
and all the models have $0.0207\,M_{\odot}$ of $^{56}{\rm Ni}$
mixed up to the mass coordinate $3.5\,M_{\odot}$ (the setup is similar to
what we later use for SN 2013ej).

In the top panel of Figure~\ref{fig:different}, we fix the density parameter
$K$ of the wind and vary its external radius. The light curves are plotted
relative to the time of shock breakout $t_0$ (as are all other light curves in
this text).
 Increasing $R_{\rm ext}$
leads to an increase of the brightness of the early light curve.
The brake in the slope
of the light curves as the luminosity falls down at
around $\sim15-30$ days coincides with the
time when the photosphere in our models passes through the
interface between the wind and the underlying RSG model.
A more extended wind effectively increases the decline rate of
the light curve, making these models promising for understanding
SNe IIL \citep{anderson:14,faran:14b,faran:14a,valenti:16}.
The values of $M_{\rm wind}$ quoted in the figure correspond to
the total mass of the wind in each case. 
The more extended winds make the total mass of the model
larger and consequently increase the length of the plateau, but this 
effect is very modest with respect to the impact on the early light curve.

In the bottom panel of Figure~\ref{fig:different}, we fix the external
radius of the wind and vary the parameter $K$. This corresponds to
everything from very low wind masses to a few especially extreme cases
where the wind is so extreme that it completely
dominates over the RSG. From this panel
one can see that there is a degeneracy in the way in which the
radius and the density of the wind impact the light curve. The more
extended and less dense winds produce light curves that are similar to
the more compact and dense winds (to see it, one can compare the
green, red and blue light curves in both panels and notice
the parameters to which these light curves correspond). This
degeneracy will be seen in Section~\ref{results}, when we
attempt to fit the observational data with the light curves from
our grid.


\section{Overview of the Supernovae}
\label{overview}

Given our result that varying $R_{\rm ext}$ seems to naturally transition from
slow to fast early declining SNe (basically, from SNe IIP to IIL), we would
next like to fit specific examples to see what properties are inferred about the
stars and their CSM environment. Due to the many parameters involved
in such fitting (e.g., $M_{\rm ZAMS}$, $E_{\rm fin}$, $R_{\rm ext}$,
$K$), this is a time consuming process. So for the present work, we focus
on three particularly well-studied events. These have been chosen for their
good multi-band light curve coverage. They also span a range of early decline
rates, with SN 2013by being the most IIL-like, SN 2013ej having an early decline
somewhat between a IIL and IIP, and SN 2013fs having a mostly flat light curve like a IIP
but also showing a particularly short plateau  (see Figure \ref{fig:2013fs_magnitudes})
that is usually observed in IIL-like objects.
This way we can see what variety of corresponding CSM properties
are inferred. Below we summarize
their main properties before fitting them in detail in Section \ref{fitting}.

\subsection{SN 2013ej}
\label{2013ej}

SN 2013ej has a moderate early decline, thus we consider it transitional between
a Type IIL and IIP.
It was discovered on 2013 July 25.45 (UT), less than 1 day after the last
non-detection, by the Lick Observatory Supernova Search
\citep[see][]{kim:13,shappee:13,valenti:13}. Details of its early photometric and
spectroscopic observations may be found in \citet{valenti:14}, 
while for the analysis of 
the pre-explosion image obtained with the {\it Hubble Space Telescope} ({\it HST})
see \citet{fraser:14}. Originally classified as Type IIP, 
this SN was reclassified later
as Type IIL, based on a fast ($1.74\,{\rm mag} / 100\,{\rm days}$ in $V$-band) 
decline rate of the luminosity as well as relatively slow 
decline of the H$\alpha$ and H$\beta$ velocity profiles, 
which are characteristic for this subclass 
\citep[see][]{bose:15,faran:14b,faran:14a}.

A  range of features observed in SN 2013ej points to a 
possible interaction of the ejecta with the CSM. 
Among them, an unusually strong absorption feature found
 in the blue wing of the H$\alpha$ P-Cygni
trough \citep[see][]{leonard:13,chugai:07}. The presence of
high-velocity components in H$\alpha$ and H$\beta$ 
profiles, demonstrated in the work of \citet{bose:15},
also suggests an interaction.
At the same time, the presence of CSM 
surrounding SN 2013ej was
supported by the X-ray measurements taken by {\it Swift} and 
{\it Chandra} instruments \citep{margutti:13}. 
\citet{chakraborti:16} analyzed these data and found
them consistent with the steady progenitor wind scenario.
According to their model, the progenitor star lost mass
at the rate $3\times 10^{-6}\,M_{\odot}\,\mathrm{yr}^{-1}$ 
assuming $v_{\rm wind}\sim10\,\mathrm{km}\,\mathrm{s}^{-1}$ 
for the last $400$ years. 

Spectropolarimetric analysis of SN 2013ej performed by 
\citet{leonard:13} revealed significant polarization of
$1.0-1.3\%$ at the early epoch ($\sim$day 7 since explosion).
Broad-band polarimetric analysis of the late ($>100$ days) 
phase of this SN performed by \citet{kumar:16} also shows 
unusually strong intrinsic polarization up to $2.14\%$. This could be
a signal of possible asymmetry in the ejecta.

SN 2013ej was previously modeled 
semi-analytically in the work of \citet{bose:15},
where its ejecta mass, radius and explosion energy were 
estimated to be $12\,M_{\odot}$, $450\,R_{\odot}$ and 
$2.3\times10^{51}\,{\rm erg}$, respectively. Hydrodynamical
simulations of \citet{huang:15} suggest an ejecta mass of $\sim10.6\,M_{\odot}$,
a radius of the progenitor of $\sim600\,R_{\odot}$, and an explosion
energy $\sim0.7\times10^{51}\,{\rm erg}$ for this SN. \citet{yuan:16}
estimate the mass of the progenitor to be $12-15\,M_{\odot}$
at ZAMS, based on the modeling of the nebular emission lines.
\cite{dhungana:16} use the approach of \citet{litvinova:83} to derive the
final pre-explosion progenitor mass of $15.2\pm4.2\,M_{\odot}$,
progenitor radius of $250\pm70\,R_{\odot}$ and explosion
energy $0.9\pm0.3\times10^{51}\,{\rm erg}$.

\subsection{SN 2013by}
\label{2013by}

SN 2013by had a particularly
steep luminosity decline ($1.46\pm0.06\,{\rm mag}$ in $V$-band) 
in the first 50 days, and we consider this representative of a IIL-like event.
It was discovered on 2013 April 23.54 (UT) by the
Backyard Observatory Supernova Search \citep{parker:13}. It was
classified as a young SN IIL/IIn  based on early optical and near
infrared observations, which was further confirmed by a detailed
analysis of \citet{valenti:15}. Possible interaction of the ejecta with the
CSM is supported by the X-ray observations 
obtained with {\it Swift} \citep[see][]{margutti:13a}. 
It also showed a very pronounced drop before transitioning to the
$^{56}{\rm Ni}$ tail, which is typical for SNe IIP. In fact, it was
shown in \citet{valenti:15} that this drop is demonstrated to a greater or lesser
extent by all of the SNe IIL that have been followed for long
enough (more than $\sim 80$ days since the discovery).

\subsection{SN 2013fs}
\label{2013fs}

SN 2013fs has a mostly flat light curve and is the most IIP-like of
all the events we consider.
It was discovered on 2013 Oct. 07.46 (UT) by Koichi Itagaki
(Teppo-cho, Yamagata, Japan) \citep{nakano:13}.
The first spectrum taken on Oct 08 with the Wide Field Spectrograph 
was reported in \citet{childress:13a} and demonstrated
extremely blue, nearly featureless continuum, exhibiting slightly
broadened emission in H$\alpha$ and H$\beta$.
This led to the preliminary classification of the object
as a SN IIn. However, the next spectrum taken on 
Oct 24 had no evidence of broadened emission and strongly
resembled a normal SN IIP spectrum \citep{childress:13b}.
Analysis of the observational data of SN 2013fs has not yet been 
presented in an individual paper, and the present work is the first
attempt to model the data numerically.


\section{Fitting the Observed Light Curves with Numerical Models}
\label{fitting}

Next we construct a grid of models and fit the multi-band light curves of
the three SNe II that are described above. We first discuss the methods
to our analysis in Section \ref{analysis} and then present results of our fitting
in Section \ref{results}. A more detailed discussion of the implications
of these fits is provided later in Section~\ref{discussion}.

\subsection{Analysis}
\label{analysis}

For our grid of models, we consider a 4-dimensional parameter space in $M_{\rm ZAMS}$,
$E_{\rm fin}$, $K$, and $R_{\rm ext}$. 
The external radius of the wind $R_{\rm ext}$ in our models varies between 
$900\,R_{\odot}$ and $2700\,R_{\odot}$ in steps of $200\,R_{\odot}$,
and $K$ takes the values 
$\{1.0,2.5,5.0,7.5\}\times10^{17}\,{\rm g}\,{\rm cm}^{-1}$, 
$\{1.0,2.5,5.0,7.5\}\times10^{18}\,{\rm g}\,{\rm cm}^{-1}$ and
$\{1.0,2.5,5.0,7.5\}\times10^{19}\,{\rm g}\,{\rm cm}^{-1}$.
We use non-rotating solar-metallicity RSG models from
the stellar evolution code \texttt{KEPLER}
\citep{weaver:78,woosley:07,woosley:15,sukhbold:14,sukhbold:16} with
ZAMS masses in the 
range between $9\,M_{\odot}$ and $16.5\,M_{\odot}$. These are
spaced in steps of 
$0.25\,M_{\odot}$ in the interval 
$9\,M_{\odot}\le M_{\rm ZAMS} \le 13\,M_{\odot}$,
and in steps of $0.5\,M_{\odot}$ in the interval 
$13\,M_{\odot} < M_{\rm ZAMS} \le 16.5\,M_{\odot}$.
The asymptotic
explosion energy $E_{\rm fin}$
(not to be confused with the inputted thermal bomb energy)
varies
in individual ranges for each SN. Namely, in the case of SN 2013ej, the
energy varies in the range $(0.4-2.6)\times10^{51}\,{\rm erg}$ for the grid
without wind, and in the range $(0.4-1.4)\times10^{51}\,{\rm erg}$ for the grid
with the wind, both in steps of $0.2\times10^{51}\,{\rm erg}$. For SN 2013by
the corresponding ranges are $(0.8-3.0)\times10^{51}\,{\rm erg}$ (without wind) and 
$(1.0-2.0)\times10^{51}\,{\rm erg}$ (with wind), and for 
SN 2013fs the ranges are $(0.4-2.6)\times10^{51}\,{\rm erg}$
and $(0.6-1.6)\times10^{51}\,{\rm erg}$. These ranges were chosen
so that the fitting parameters are located well inside the grid to maximize
computing resources.

In addition to the grid, for each SN we use a fixed mass of radioactive nickel $M_{\rm Ni}$,
which was taken from the supporting
information in \citet{valenti:16}.
This gives $M_{\rm Ni}=0.0207\pm0.0019$ for SN 2013ej,
$0.032\pm0.0043$ for SN 2013by, and 
$0.0545\pm0.0003$ for SN 2013fs. In all models we mix the $^{56}{\rm Ni}$
up to the mass coordinate $3.5\,M_{\odot}$ and do not vary this parameter.

Before comparing the numerical models with the data, we
corrected the multi-band light curves for
reddening using the Cardelli law \citep[see][]{cardelli:89}. The following values for the
absorption in $B$-band were used: $A_{B} = 0.25\,{\rm mag}$ for 
SN 2013ej, $A_{B} = 0.798\,{\rm mag}$ for 
SN 2013by and $A_{B} = 0.145\,{\rm mag}$ for SN 2013fs. 
In all three SNe, the reddening is due to the Milky Way, while the
contribution from the host galaxy is negligible. The distance 
moduli are $\mu=29.79\pm0.02$ for SN 2013ej, $\mu=30.84\pm0.15$ for 
SN 2013by and $\mu=33.50\pm0.15$ for SN 2013fs \citep{valenti:16}.

For each of the three SNe we assess the best fitting model
within the generated grids of light curves by calculating $\chi^2$ as
\begin{equation}
\chi^2 = \sum_{\lambda\in[g,...,z]} \,\,\, 
\sum_{t^*<t_{PT}}\frac{(M_{\lambda}^*(t^*)-M_{\lambda}(t^*))^2}{M_{\lambda}(t^*)}\ ,
\end{equation}
where $M_{\lambda}^*(t^*)$ is the observed magnitude in
a given band $\lambda$ at the moment of observation $t^*$,
$M_{\lambda}(t^*)$ is the numerically obtained magnitude
in the same band at the same moment of time, and $t_{PT}$ is the
length of plateau as defined in the work of \citet{valenti:16}. This is
taken to be
$99\,{\rm days}$ for SN~2013ej, $85\,{\rm days}$ for SN~2013by
and $83\,{\rm days}$ for SN~2013fs. We restrict our fits to the
plateau phase, because during the radioactive tail we 
do not expect the spectrum to be
well described by a black body, as assumed in \texttt{SNEC}.
The best
fitting model corresponds to the minimum value of $\chi^2$.
We include in the $\chi^2$ all bands redder than $g$-band and do
not include $B$-, $u$- and $U$-bands, since the light curves
in these bands are affected by iron group line blanketing \citep{kasen:09},
which is not taken into account in \texttt{SNEC}.

When calculating $\chi^2$, we do not take into account the
observational errors, because their values may differ by an order
of magnitude for different epochs. 
Taking into account the error bars leads to a visually worse fit, because the best
fitting model tends to fit a few particular points with the least error instead
of fitting the entire light curve.

\begin{figure}[!h]
  \centering
  \includegraphics[width=0.475\textwidth]{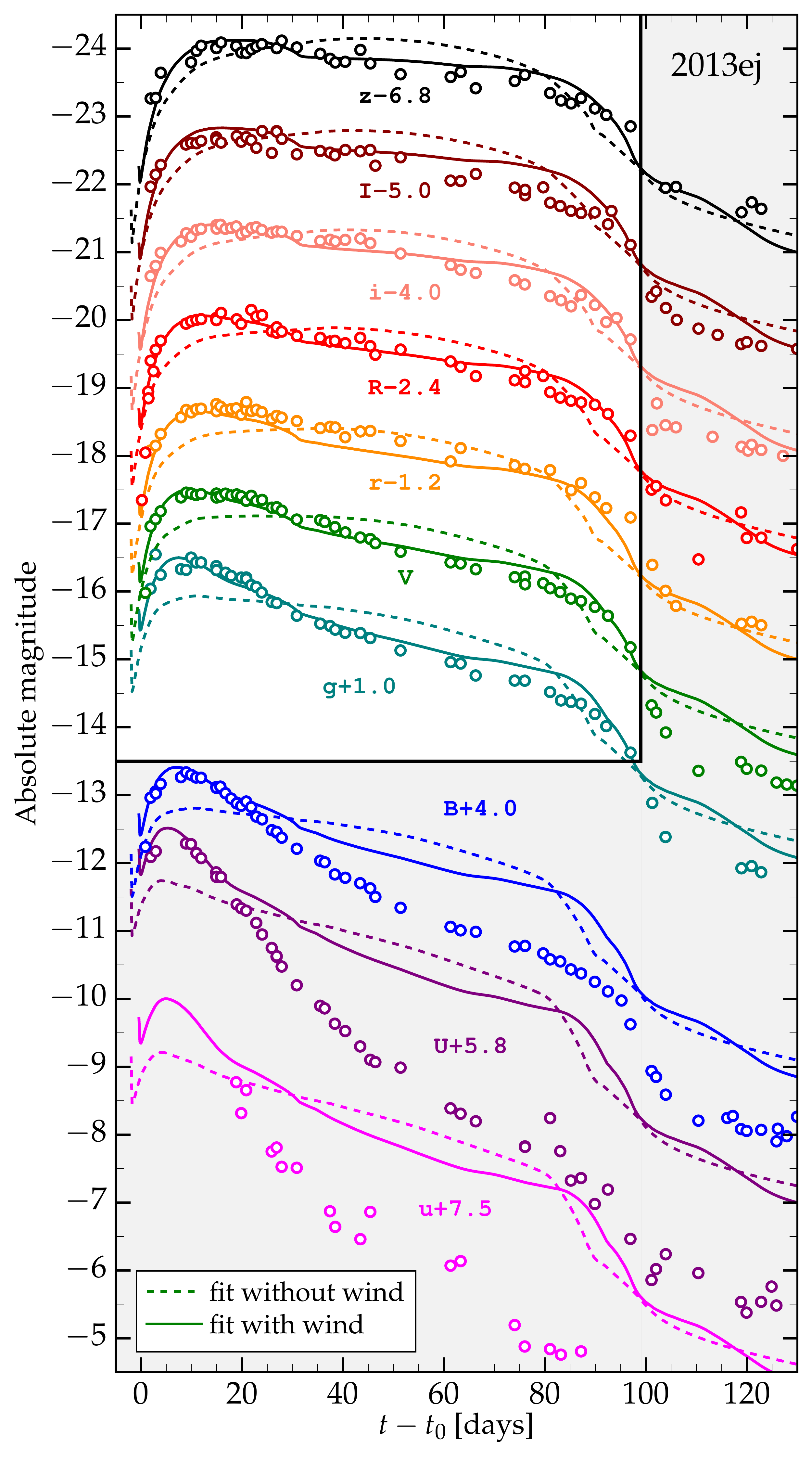}
  \caption{Best fits of the SN 2013ej data without (dashed lines) and
  with (solid lines) CSM.} 
  \label{fig:2013ej_magnitudes}
\end{figure}
%


\subsection{Fitting Results}
\label{results}

\begin{figure}[!h]
  \centering
  \includegraphics[width=0.475\textwidth]{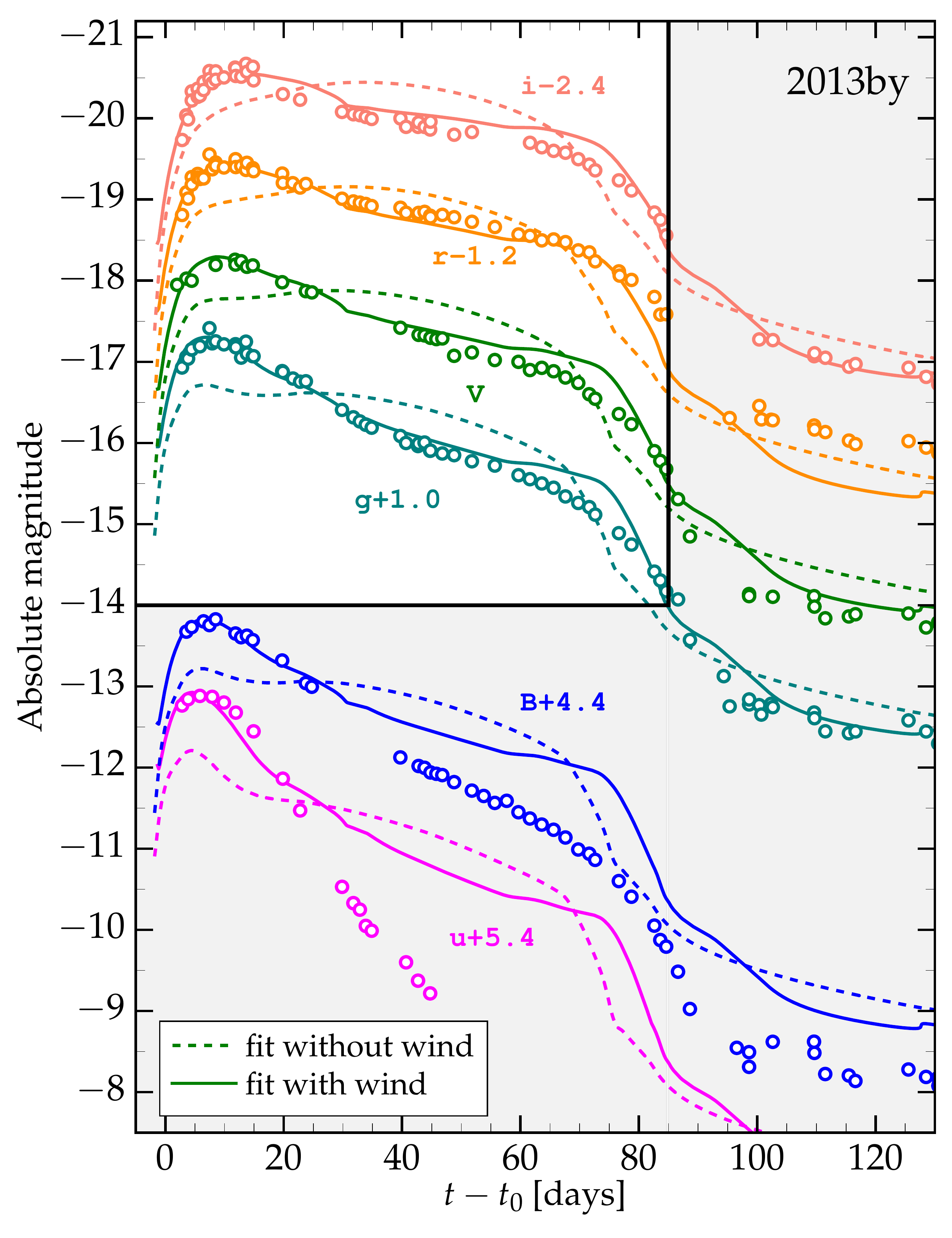}
  \caption{Best fits of the SN 2013by data without (dashed lines) and
  with (solid lines) CSM.} 
  \label{fig:2013by_magnitudes}
\end{figure}

\begin{figure}[!h]
  \centering
  \includegraphics[width=0.475\textwidth]{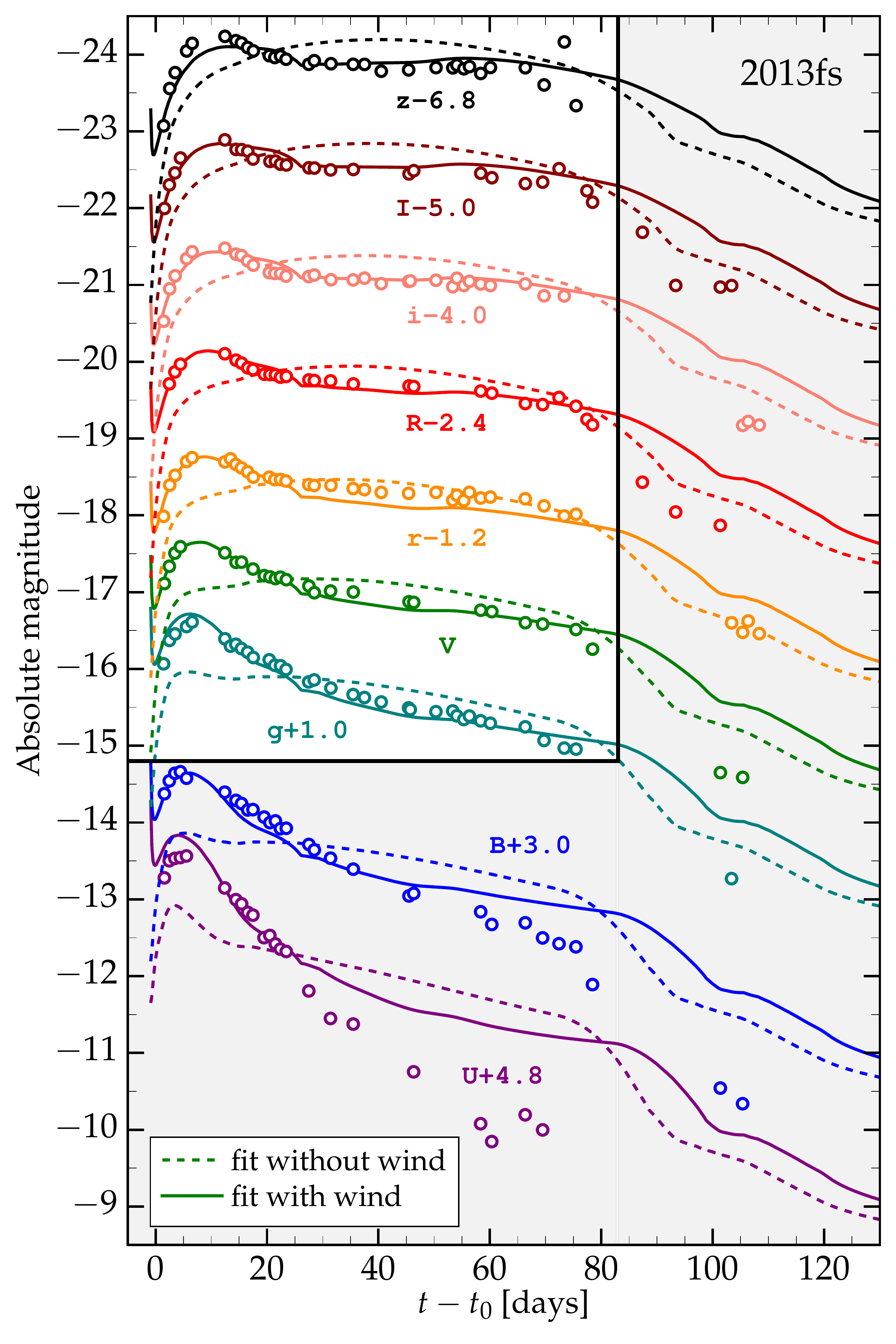}
  \caption{Best fits of the SN 2013fs data without (dashed lines) and
  with (solid lines) CSM.} 
  \label{fig:2013fs_magnitudes}
\end{figure}

Figures~\ref{fig:2013ej_magnitudes} to \ref{fig:2013fs_magnitudes} 
show the data in different bands for the three
considered SNe, together with our best fitting models with (solid lines) and without (dashed lines)
CSM. The white (unshaded) regions in the plots contain the
data which was used in order to calculate the fits, while the data from the
gray (shaded) regions was not used in our analysis. Comparing the
dashed lines with the data in Figures~\ref{fig:2013ej_magnitudes} to \ref{fig:2013fs_magnitudes} 
demonstrate that none of the light curves obtained from the RSG
models without wind can reproduce the data well. Without a wind,
the light curves are not sufficiently peaked at early times. The fitting routine
compensates for this by ``splitting the difference'' and overshooting the data
during the plateau phase. In contrast, the models that include a
wind provide a much better fit across all the data.


It is interesting to note that our models with the wind
even reproduce reasonably well the early parts of the $u$-,
$U$- and $B$-data, which we did not explicitly used to find the fit. Large
discrepancies between our light curves and the data in these
bands at later times can be explained by the iron group line
blanketing, which is not taken into account in \texttt{SNEC}. In
the work of \citet{kasen:09}, it was shown that this effect starts playing 
an important role for the blue bands after few tens of days. Similarly,
our results suggest that for the first $\sim10-20$ days the 
effect of the iron group line blanketing is not so strong, and the
spectrum can be well described by a black body.
The transition between the plateau and $^{\rm 56}$Ni tail is sensitive to the
low temperature opacities, which are not well known, as well as
to the degree of mixing of $^{\rm 56}$Ni, which we did not vary
in this study because it would just be too many parameters to fit.
So we do not view places where we are not able to reproduce the data during this
phase as a failure of the model.

\begin{figure*}
\begin{center}
\includegraphics[width=0.33\textwidth]{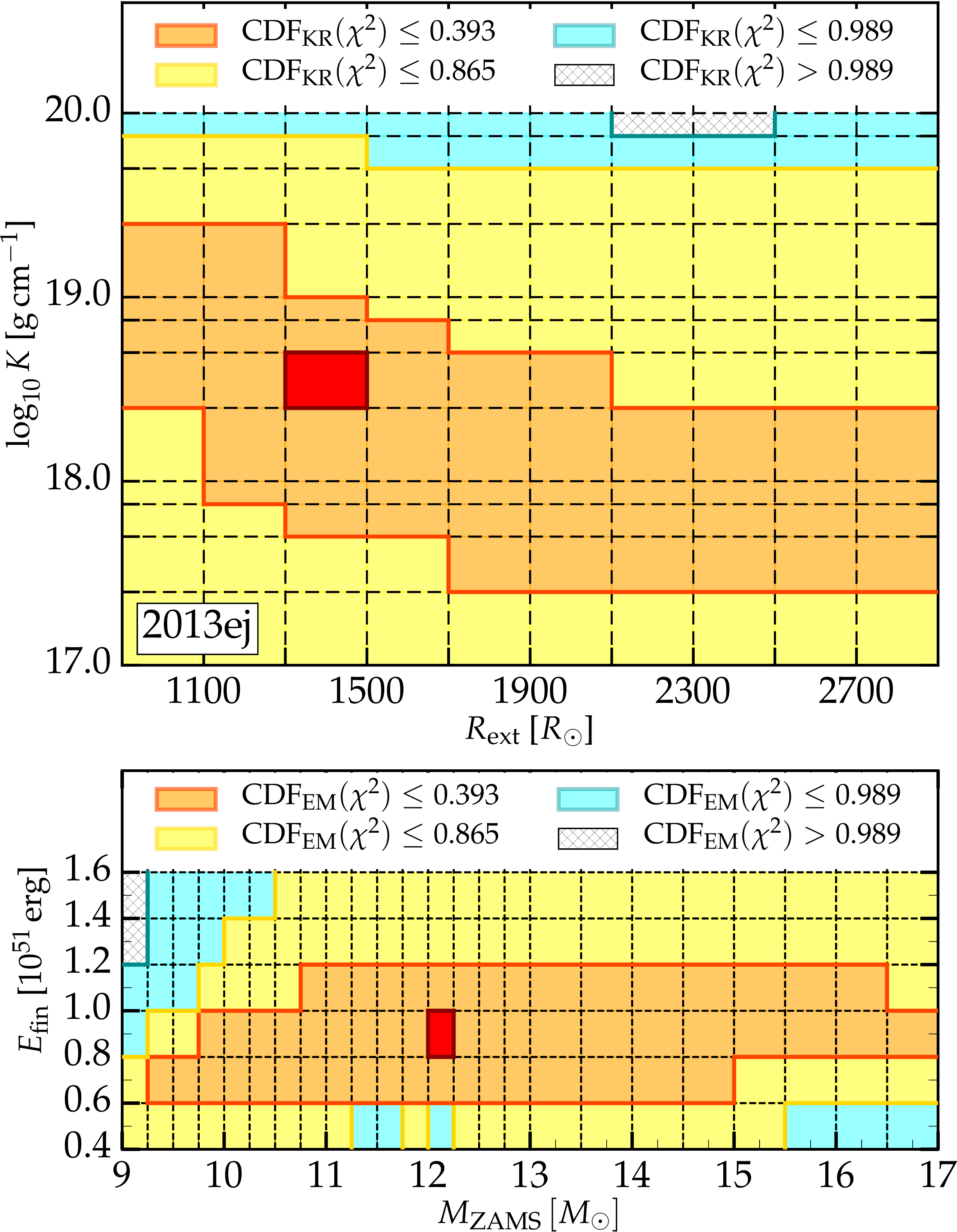}
\includegraphics[width=0.33\textwidth]{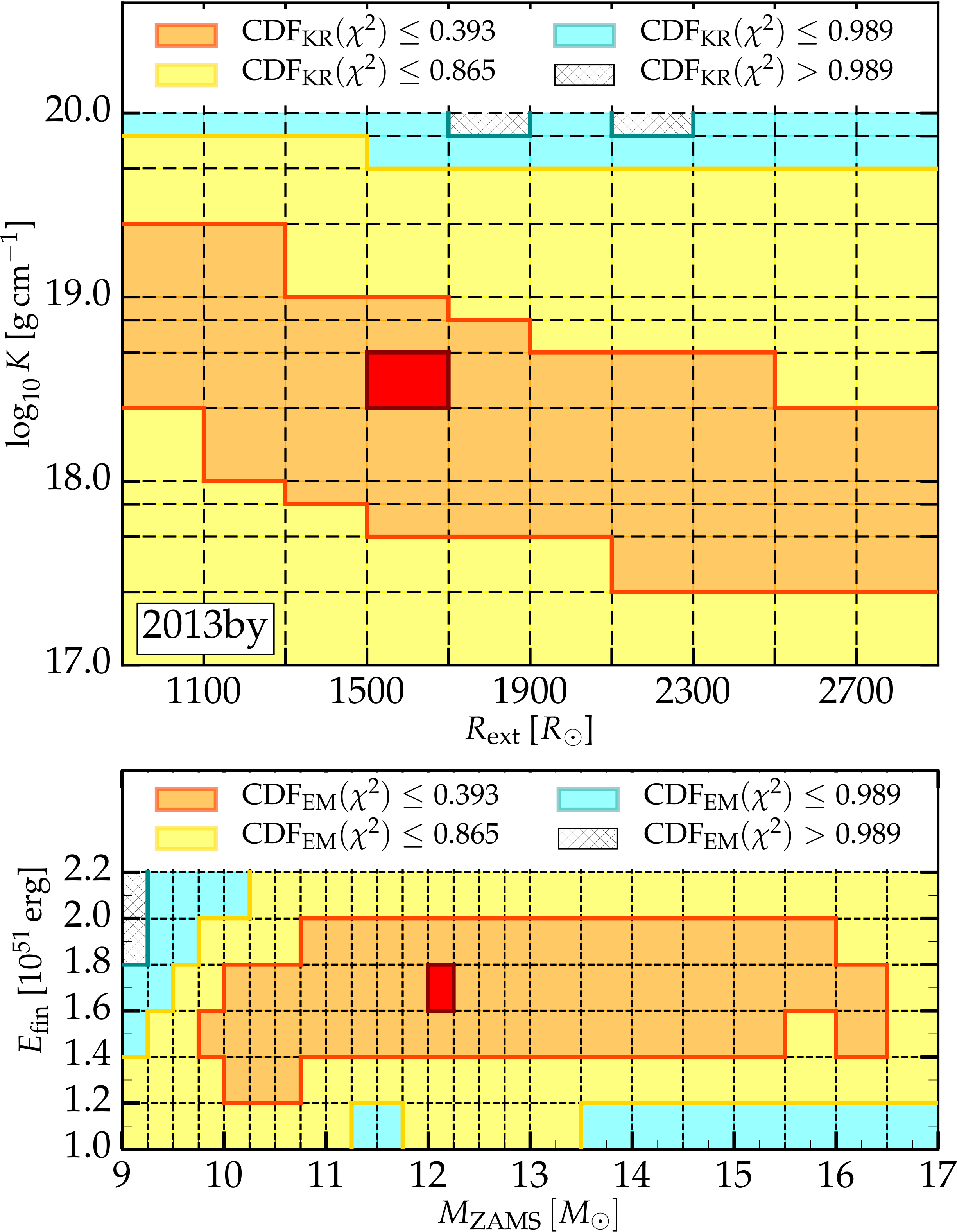}
\includegraphics[width=0.33\textwidth]{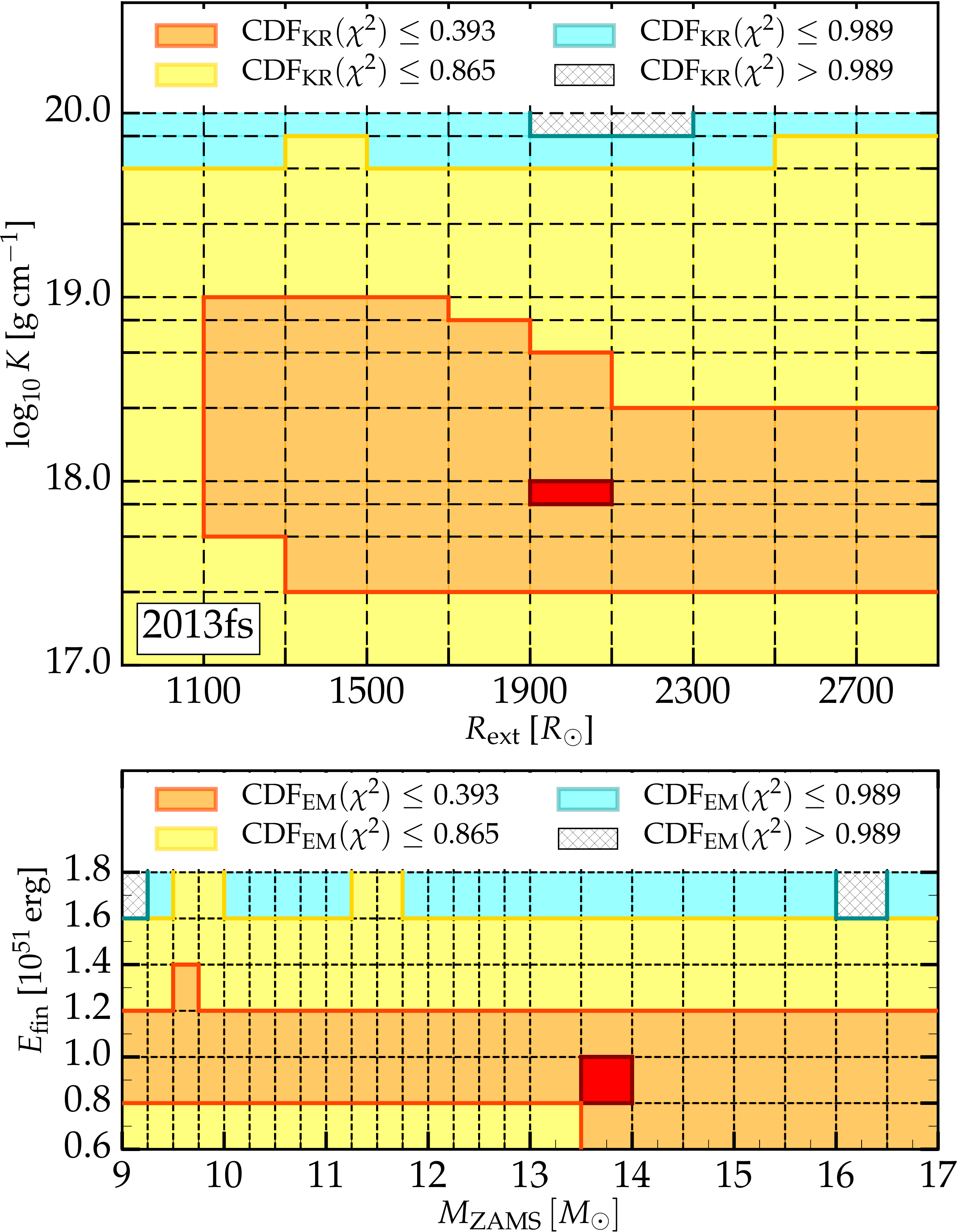}
\end{center}
\caption{Top panels: 2D slices ($E_{\rm fin} = E_{\rm fin,fit}$,
  $M_{\rm ZAMS}=M_{\rm ZAMS,fit}$) of the 4D 
   parameter space for SN 2013ej, SN 2013by and SN 2013fs.
   Bottom panels: 2D slices
  ($R_{\rm ext}=R_{\rm ext,fit}$, $K=K_{\rm fit}$) of the 4D 
  parameter space for SN 2013ej, SN 2013by and SN 2013fs.
  Confidence regions correspond to
  the $1$, $2$ and $3$ standard deviations of the mean of a 
  two-dimensional symmetric Gaussian (see text for additional details).} \label{fig:chisq}
\end{figure*}

Figure~\ref{fig:chisq} shows the position of the fitting 
parameters for the SNe 2013ej, 2013by and 2013fs. 
The top panels of the figure show the $K-R_{\rm ext}$
slice of the 4D parameter space, where the values of 
$M_{\rm ZAMS}$ and $E_{\rm fin}$ are set
equal to the fitting values. The bottom panels of the figure
shows the $M_{\rm ZAMS}-E_{\rm fin}$ slice of the 4D parameter space, 
where the values of $R_{\rm ext}$ and $K$ are set equal to the fitting values. 
Red blocks in all the plots denote the best fitting parameters, which
 we summarize in Table~\ref{tab:parameters}.
To evaluate the robustness of the fits,
we show the $39.3\%$, $86.5\%$ and $98.9\%$ confidence regions,
which correspond to the one, two and three standard deviations
of the mean of a two-dimensional Gaussian\footnote{The values of
$0.393$, $0.865$ and $0.989$ may be obtained by solving the integral 
$\int_0^{2\pi}d\phi\int_r r (2\pi \sigma^2)^{-1} \exp 
\left(-r^2/2\sigma^2\right)dr$ in the regions $0\le r\le\sigma$, 
$0\le r\le 2\sigma$, and $0\le r\le 3\sigma$, respectively 
\citep[see][]{andrae:10}. Here
$\sigma$ is the standard deviation of a two-dimensional symmetric 
Gaussian given in polar coordinates $(r,\phi)$.}.

One can see from Figure~\ref{fig:chisq}
that there are strong degeneracies in some of the parameters.
The confidence interval in the $K-R_{\rm ext}$ plane has a 
characteristic ``banana'' shape, in the sense that the models with 
extended low density wind produce similar fits to the models with
less extended high density wind. Also, $M_{\rm ZAMS}$ is difficult to
completely constrain because a large part of the light curve we are fitting
is hidden by the CSM. This is not surprising since \citet{kasen:09} show
that the length of the plateau is only weakly dependent on the ejecta mass.
Nevertheless, it is robust that {\it light curves with a dense wind do
a dramatically better job at fitting the data than those without}.

It is also interesting to compare our best fit parameters for SN~2013ej to
the results of \citet{nagy:16}, who use a semi-analytic, two-component
model to fit the same SN. They find a mass and radius for what they call
the ``envelope'' material
of $0.6\,M_\odot$ and $980\,R_\odot$, respectively, while for our wind we
find $0.7\,M_\odot$ and $1300\,R_\odot$. Given the difference in techniques
(they focus on fitting bolometric light curves while we are fitting various
photometric bands with numerical models), the similarity of these inferred
parameters is encouraging. The fact that \citet{nagy:16} extend this sort of
fitting to a variety of other well-studied Type IIP-like SNe argues that
a similar amount of CSM as we infer for SN~2013ej may be present
in a wide range of otherwise seemingly ``normal'' events.

\begin{table*}[!ht]
\renewcommand{\arraystretch}{1.3}
\caption{Best Fit Parameters}
    \label{tab:parameters}
\centering
    \begin{tabular}{ccccccc}\hline\hline
      Name & $M_{\rm ZAMS,fit}\,[M_{\odot}]$ & $E_{\rm fin,fit}\,[10^{51}\,{\rm erg}]$ 
      & $K_{\rm fit}\,[{\rm g}\,{\rm cm}^{-1}]$ & $R_{\rm ext,fit}\,[R_{\odot}]$ & 
      $\dot{M}\footnotemark[1]\,[M_{\odot}\,{\rm yr}^{-1}]$ & $t_{\rm wind}\footnotemark[1]$     \\\hline
      SN 2013ej & 12.0 & 0.8 & $2.5\times10^{18}$ & 1300 & 0.5 (5.0) & $1.5\,{\rm yr}$ ($1.8\,{\rm months}$) \\
      SN 2013by & 12.0 & 1.6 & $2.5\times10^{18}$ & 1500 & 0.5 (5.0) & $1.9\,{\rm yr}$ ($2.3\,{\rm months}$) \\
      SN 2013fs  & 13.5 & 0.8 & $7.5\times10^{17}$ & 1900 & 0.15 (1.5) & $2.5\,{\rm yr}$ ($3.0\,{\rm months}$) \\
\hline
    \end{tabular}
       \footnotetext[1]{Both $\dot{M}$ and $t_{\rm wind}$ are estimated using a wind velocity $v_{\rm wind}=10\,{\rm km\,s^{-1}}$
       ($v_{\rm wind}=100\,{\rm km\,s^{-1}}$).}  
\end{table*}

\section{Discussion}
\label{discussion}

Next we discuss some of the general trends that are seen in these fits,
the physics that produces these features, and the
implications for the exploding progenitors given the CSM properties we infer. This
includes both what the progenitors look like and what physical processes may
have caused them to be this way.

\subsection{Light Curve Properties}

\begin{figure}[h!]
  \centering
  \includegraphics[width=0.475\textwidth]{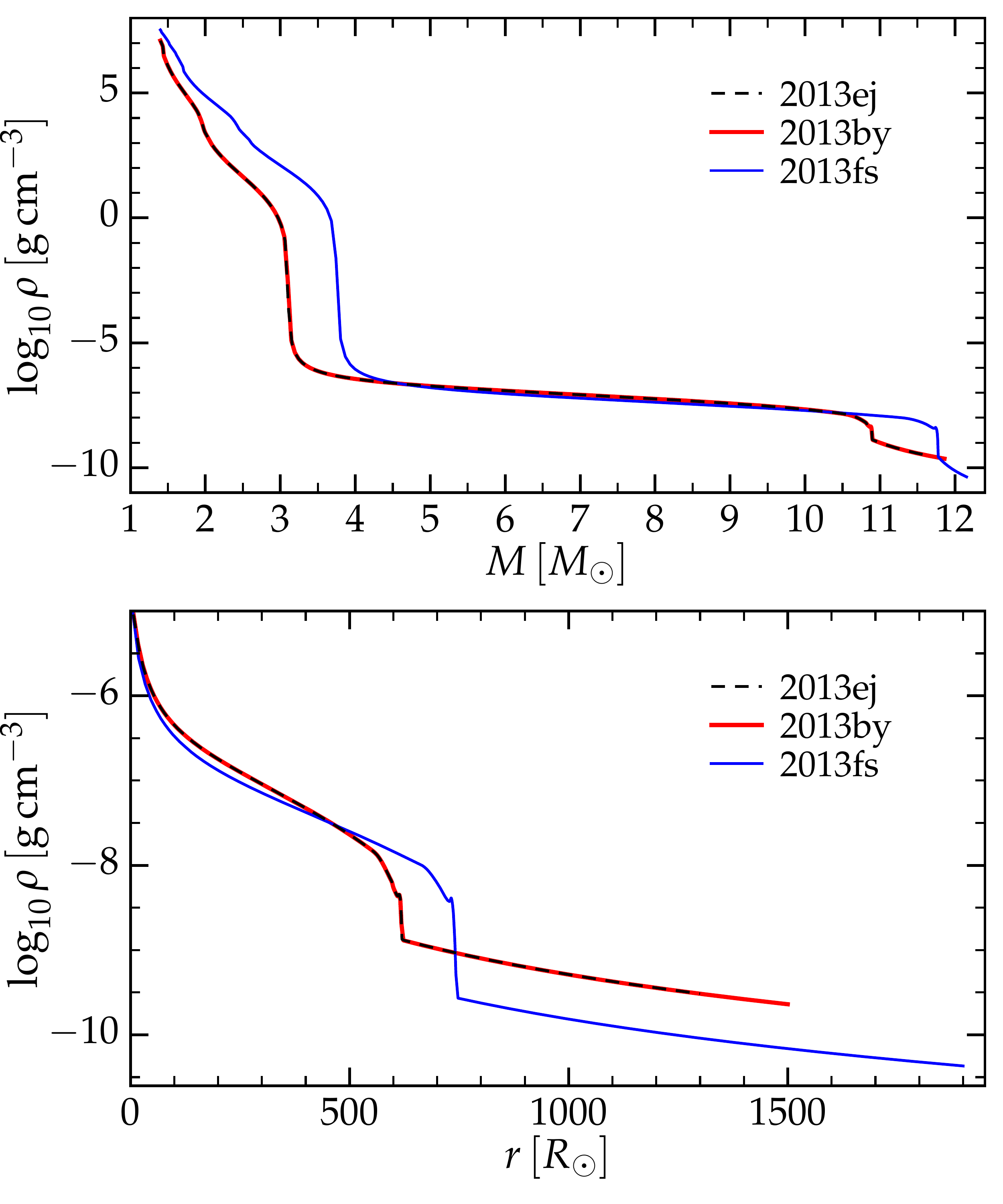}
  \caption{Density profiles of the best fitting models as a function 
  of mass coordinate (top panel) and radial coordinate (bottom panel)
  as summarized in Table~\ref{tab:parameters}. The
  progenitor models for SN 2013ej and SN 2013by differ only by 
  $R_{\rm ext}$.} 
  \label{fig:profiles}
\end{figure}

Figure~\ref{fig:profiles} shows the density profiles of the best fitting models 
for SNe~2013ej, 2013by and 2013fs as a function 
of mass (top panel) and radius (bottom panel). They are quite
similar to each other, with the models for SNe~2013ej and
2013by differing only by the external radius of the wind.
In the wind picture, this would just correspond to a similar $\dot{M}$
but a different amount of time between the start of the wind and explosion,
in this case a difference of $\sim100\,$days depending on the
wind velocity.
To understand how this impacts the light curve,
one can draw an analogy between these models 
and the extended envelopes considered in the context of double-peaked 
SNe IIb \citep{nakar:14,piro:15}. Here, instead of the low density extended material
attached to a compact core, we have the wind surrounding
higher density RSG models. 
For this kind of progenitors, the initial brightness and fast rise of the light
curve are explained by the cooling emission from the low density
material (wind), which does not experience strong adiabatic losses due to its
initially large volume. The maximum in the optical bands
for these progenitors correspond to the moment, when the luminosity shell 
(the depth from which photons diffuse to reach the 
photosphere at a given time after shock breakout) reaches the
interface between the low density and the high density material \citep[see discussion in][]{nakar:14}.

\begin{figure}[h!]
  \centering
  \includegraphics[width=0.475\textwidth]{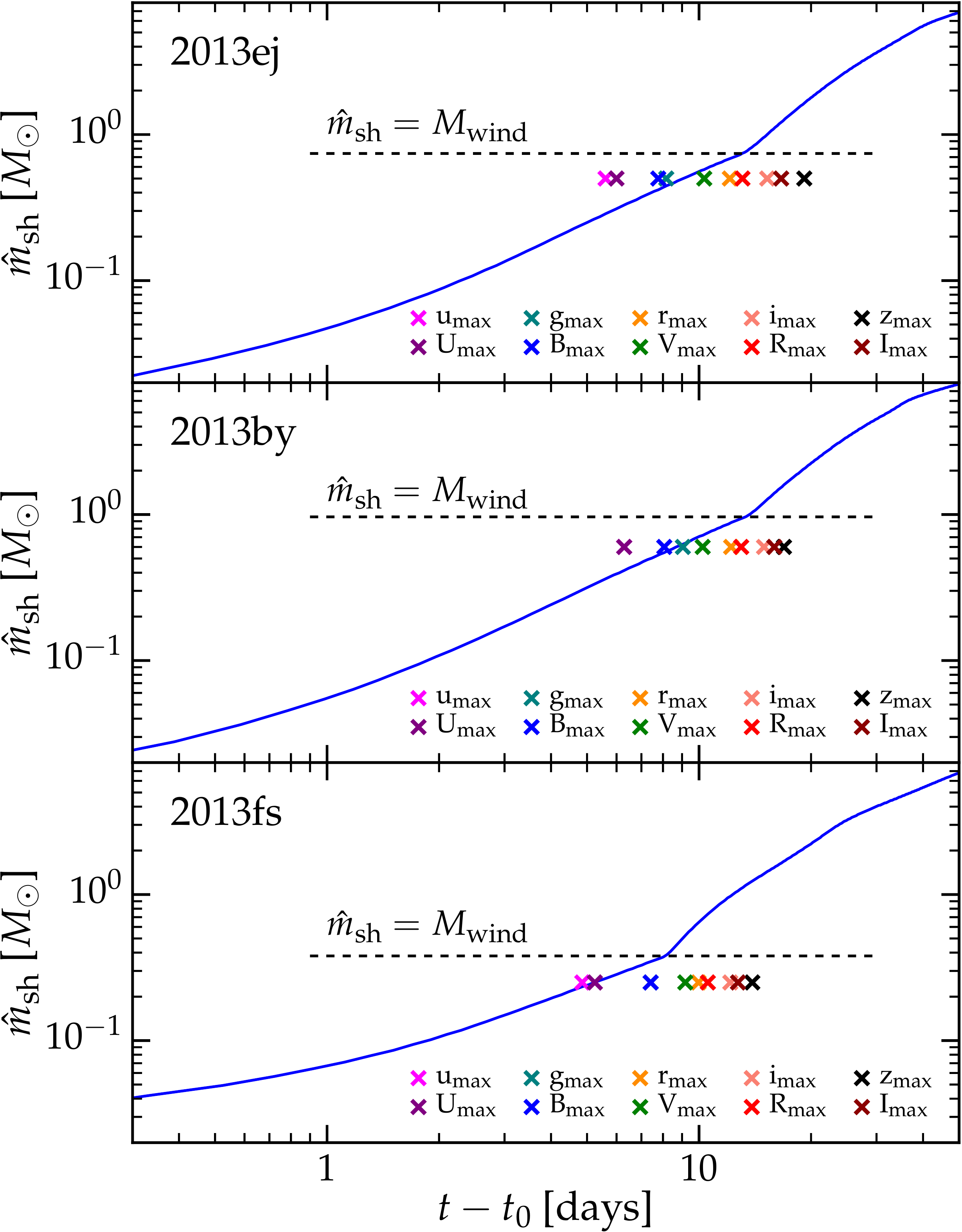}
  \caption{$\hat{m}_{\rm sh}$, the difference between the
    total mass of the model and the mass coordinate of the luminosity
    shell, as a function of time since shock breakout ($t_0$) for the best
    fitting models of SN 2013ej, SN 2013by and SN 2013fs. Crosses
    indicate the times of maxima in the corresponding bands. Maxima in
    the optical bands approximately coincide with the moment when
    the luminosity shell crosses the interface between the RSG model 
    and the wind, $\hat{m}_{\rm sh} = M_{\rm wind}$.} \label{fig:lumshell}
\end{figure}

To illustrate this, Figure~\ref{fig:lumshell} shows the time dependence of $\hat{m}_{\rm sh}$,
which is defined as the difference between the total mass of the
progenitor and the mass coordinate of the luminosity shell, for the three
fitting models. The position of the
luminosity shell is found from the condition $\hat{t}_{\rm diff} = t-t_0$, 
where $t_{\rm diff}=t_{\rm diff}(r,t)$ is the diffusion time at each
moment of time and at each depth, and the hat
indicates the value of this quantity taken specifically at the
luminosity shell \citep{nakar:10}. The diffusion time is computed using
\be
	\hat{t}_{\rm diff} = \int_{\hat{r}}^{R_{\rm ext}} \frac{3\tau dr}{c},
\ee
where
\be
	\tau(r) = \int_r^{R_{\rm ext}} \kappa \rho dr,
\ee
and $\hat{r}$ is the radius of the diffusion depth \citep[see][for more details]{morozova:16}.
From Figure~\ref{fig:lumshell} it is clear that the slope of 
$\hat{m}_{\rm sh}$ as a function of time changes abruptly when the
luminosity shell passes the interface between the wind and the underlying 
RSG model, $\hat{m}_{\rm sh}=M_{\rm wind}$. At the same time, the
light curves in the optical bands pass through their maxima
shown by the crosses in the plots

\subsection{The Origin of the SN IIL Classification}

The results of our fitting argue that the origin of SNe IIL
is the addition of dense CSM, relatively closely position to the star. 
This is in contrast to a number of previous studies that explored
reducing the mass of the progenitor in order
to get SN~IIL-like light curves \citep{blinnikov:93,morozova:15,moriya:16}, but similar to the
suggestion in \citet{smith:15} from the observation of PTF11iqb.
Even beyond our modeling though, all three of the considered SNe
demonstrated some signs of CSM interaction in the observations,
as we emphasize in Section \ref{overview}, strengthening our
conclusions. In this picture, continuous transition between
the SNe IIP and IIL \citep{anderson:14,sanders:15}
is naturally explained by the continuity
in the wind properties of the progenitors.
Also, this approach naturally explains
observed positive correlation between the maximum 
brightness and the decline rate of SNe \citep{anderson:14}, because
adding the wind to the progenitor profile increases both of these
two observables. We note that the longstanding tradition of presenting groups of
SNe IIP and IIL aligned at maximum brightness may have been somewhat misleading
because it suggested that the SNe IIL had something missing when in fact they
have something added in comparison to a typical RSG.

The question of whether all SNe IIL demonstrate (at least 
moderate) interaction with CSM 
has been already raised in \citet{valenti:15}
and discussed in \citet{bose:15}, but our work highlights
even more how important this is to investigate.
In some cases, these interactions appear through
narrow lines seen at early times \citep{gal-yam:14,smith:15,khazov:16}.
For the spherically symmetric calculations we perform here, the wind
is always optically thick and may appear at odds with these observations.
This can be reconciled though if there is additional low density material above
the wind that we are calculating.
Another possibility is if the CSM is non-spherical, so that
the component we are calculating represents regions where the shock
can pass into the CSM. In such a case, narrow lines may be formed
by tenuous material above and below the CSM, and would therefore disappear
once the ejecta overtakes the CSM within a matter of $\lesssim2\,$days
(given the radii we are inferring for it). {\it Therefore
to see spectral signatures of the CSM requires getting spectra as soon after
explosion as possible.} Furthermore, depending on the geometry of the CSM, not all of it may be
disrupted by the explosion, and some may still be present once the ejecta has passed.
This has in fact been seen for PTF11iqb, for which it was argued that the
CSM may be in the form of a disk \citep{smith:15}. Thus it may also be helpful to
perform late observations on SNe with a fast early decline to identify how ubiquitous such features are.

\subsection{Implications for the Stellar Progenitors of SNe IIL}

Up to now, pre-explosion imaging has not revealed any clear
distinction between the progenitors of Type IIP and Type IIL SNe
neither in radius \citep{gall:15} nor in mass \citep{valenti:16}.
Since our winds are optically thick, the radius we are finding $R_{\rm ext}$ would
act as a ``pseudo-photosphere,'' making the progenitors much redder and
potentially dimmer if dust formation occurs.
This may appear to be at odds with the pre-explosion imaging. Unfortunately,
most pre-explosion imaging is not close enough to
time of explosion to probe this CSM given the timescales we are inferring
(see Table \ref{tab:parameters}). For example, SN 2009kr was a IIL
inferred to have a reasonably normal progenitor radius of $\sim400-800\,R_\odot$,
which is consistent with the radii of standard RSGs
\citep{levesque:05,levesque:06}. Unfortunately,
 this was done with {\it HST} imaging that occurred
between $\sim10$ and $\sim2\,$yrs before explosion \citep{fraser:10,elias-rosa:11}.
In another case, SN 2013ej (which is one of the events we studied here) was inferred
to have a radius  $\sim400-800\,R_\odot$ \citep{valenti:14,fraser:14}, but again
this is for {\it HST} imaging between $\sim10$ to $8\,$yrs before the SN in
comparison to the $\lesssim2\,$yrs we infer for this event.

Our results highlight that pre-explosion imaging will be needed within
$\sim\,$yr of explosion to provide meaningful constraints on the CSM environment
that may be turning these events into SNe IIL. Furthermore, given the large
radii we infer, these progenitors should be extremely red if not also
extinguished by dust. This would make them  look anomalously dim
at optical wavelengths so that they are mistaken for lower mass stars unless
infrared coverage is available.

\subsection{Source of the Circumstellar Material}

The mass loss rates quoted in Table~\ref{tab:parameters}
are much higher than the ones observed for 
steady winds in RSGs \citep{nieuwenhuijzen:90}. 
Even for the extremely dense wind of the most luminous known RSG
\citep[VY CMa,][]{smith:09}
the estimated mass loss rate is
$1-2\times 10^{-3}\,M_{\odot}\,{\rm yr}^{-1}$.
At the same time,
analysis of the emission lines of Type IIn SNe suggests rather
strong winds before the explosion
\citep[see][but note higher wind velocities than ours]{kiewe:12,smith:14}.
This suggests that the CSM we are modeling as a wind may not be
a wind at all but represent a more explosive outburst rather than
something steady.

There is increasing evidence that violent outbursts are
important in the final stages of the lives of massive stars. This evidence
can come in the form of direct detections of pre-SN outbursts
\citep[e.g.,][]{foley:07,pastorello:07,smith:10,foley:11,mauerhan:13,pastorello:13}
or inferred from the dense CSM needed to explain SN light curves
\citep[e.g.,][]{smith:07,smith:07b,ofek:07,smith:08,fransson:14}.
Although how common such eruptions are is still being investigated
\citep[see][]{ofek:14,bilinski:15}, our work may indicate
that SNe IIL are just lower mass versions of these.
It is interesting to note that the time of the wind
$t_{\rm wind}$ given 
in Table~\ref{tab:parameters} is comparable
to the duration of the carbon shell burning for an
$M_{\rm ZAMS}=12\,M_{\odot}$ star \citep[see][]{fuller:15}. 
Increasing velocity of the wind 
by a factor of $10$ would boost the mass loss rates
to even larger values, 
but in this case $t_{\rm wind}$ would coincide with the duration of the
oxygen shell burning ($\sim$few months).
A better measurement of the extent of the CSM along with its velocity may
help to better connect its properties with various stages of stellar
burning. This will assist in identifying its physical origin
\citep[see discussions in][and references therein]{smith:arnett:14,woosley:15,quataert:16}.

In these explosive scenarios, the CSM is likely not to have the density profile
of a steady wind as we have assumed. In such cases, the radius and mass
we infer is probably just an estimate for the true CSM properties. There may
also be other possibilities. For example, the CSM we are inferring could be
an inflated outer radius of the star, perhaps driven by additional energy
input during the end of the star's life \citep[e.g., like from waves,][]{quataert:12,shiode:14}.
Another possibility is that the material could be in the form of a disk as
discussed above for PTF11iqb \citep{smith:15}. In any case, better pre-explosion
imaging, early observations during the first $\sim1\,$day after explosion, and the theoretical
modeling of various CSM distributions should help piece together the true
properties of the CSM surrounding SNe IIL.

\section{Conclusion}
\label{conclusion}

We have numerically investigated the light curves of RSGs with CSM. These
vary from most previous theoretical studies in that the winds we consider
are generally more dense and
are rather compact, only extending a few stellar radii above the RSG.
We found that the corresponding light curves show many of the hallmarks of
SNe IIL and then fit the observations of three particular well-studied SNe II with
RSG plus CSM models. The key inferred properties of the CSM are mass loss rates
of $\sim0.1-0.5\,M_\odot\,{\rm yr}^{-1}$ and an extent of $\sim1300-1900\,R_\odot$,
which implies that the CSM was generated $\lesssim1-2\,$yrs prior to explosion.
This may indicate that this material may be driven by certain advanced stages
of stellar burning, but since these estimates  depend on the uncertain velocity of the CSM,
it is possible that it may be occurring on shorter timescales.

Our results highlight that pre-explosion imaging $\lesssim1\,$yr prior to explosion
and spectra taken $\lesssim2\,$days following explosion will be key for investigating
the properties of this CSM. There should be trends between the early time light curve
slope and the inferred CSM properties, and these need to be explored to build a more
complete picture on the nature of the CSM. In other cases, one might expect signatures
of the CSM to pop up at later times like PTF11iqb \citep{smith:15}, depending on its radial
and latitudinal
distribution. Surveys with rapid cadences, such as the Zwicky Transient Facility \citep{law:09}
and the All-sky Automated Survey for Supernovae \citep{shappee:14} make this an ideal time to
identify SNe early, so that these critical spectra can be taken.
The Large Synoptic Survey Telescope \citep{lsst} could be also
useful in this respect depending on its final cadence. Nevertheless, LSST
could also be helpful for having an archive of
good time coverage for these progenitors before they explode. This way after the SNe are
discovered, their history can
be investigated to see whether they showed any pre-explosion outbursts or enhanced winds 
as we infer here.

\acknowledgments
We acknowledge helpful discussions with and feedback from 
A.~Burrows, D.~Clausen, S.~M.~Couch, J.~Fuller, D.~Milisavljevic, C.~D.~Ott,
D.~Radice, B.~J.~Shappee, N.~Smith, T.~Sukhbold, and J.~C.~Wheeler.
This work is supported in part by the National Science Foundation
under award Nos.\ AST-1205732 and AST-1212170, by Caltech, and by the
Sherman Fairchild Foundation. The computations were
performed on the Caltech compute cluster Zwicky (NSF MRI-R2 award
no.\ PHY-0960291) and on the MIES cluster of the Carnegie Observatories, which was
made possible by a grant from the Ahmanson Foundation.

\bibliographystyle{apj}

\end{document}